\documentclass[a4paper,fleqn]{cas-sc}

\usepackage[numbers]{natbib}

\usepackage{placeins,quoting}

\newcommand\fref[1]{\normalsize{\textbf{\textit{Figure \ref{#1}}}}\normalsize{}}                                                                                                    
\def\tsc#1{\csdef{#1}{\textsc{\lowercase{#1}}\xspace}}
\tsc{WGM}
\tsc{QE}
\tsc{EP}
\tsc{PMS}
\tsc{BEC}
\tsc{DE}

\begin{document}
\let\WriteBookmarks\relax
\def\floatpagepagefraction{1}
\def\textpagefraction{.001}
\shorttitle{Race and Running} 
\shortauthors{Souaiaia et~al.}

\title [mode = title]{Revisiting Stereotypes: Race and Running}

\author[1]{Tade Souaiaia}[orcid=0000-0003-3922-1372] 
\cormark[1]
\ead{tade.souaiaia@downstate.edu} 
\credit{conceptualisation, investigation, formal analysis, methodology, writing -original draft}
\address[1]{Department of Cell Biology, SUNY Downstate Health Sciences, Brooklyn, NY, USA} 
\cortext[cor2]{Corresponding author}

\author[2]{Nabie Fofanah}[]
\address[2]{Olympic Athlete (100m), Professional Speed Coach, SD88 LLC}
\credit{conceptuatlisation, investigation}

\author[3]{Rawle DeLisle}[]
\address[3]{USATF Level 2 Sprints and Hurdles Coach, Ballistic Sprint LLC}
\credit{data curation, investigation, writing -review \& editing}

\author[4]{Sheena Mason}
\address[4]{Department of English, SUNY Oneonta, Oneonta, NY, USA} 
\credit{investigation, writing -original draft}

\begin{abstract}
The athletic achievements of African athletes in global running championships have long been subject to 
scientific and sociological inquiry. During the 1990s, a popular narrative emerged, suggesting that
West African lineage conferred inherent sprinting advantages, and that North, South and East African's 
are specialized for longer distances. Part and parcel to this narrative was the enthusiastic belief that 
it would \textit{very soon} be substantiated by a genotyping revolution that would enable prognostication 
of individual athletic potential.

\noindent We revisit this hypothesis in the post-genomic era.  First, we compare the global running records 
used to generate the racialist hypotheses with performances over the last twenty years (2004-2023).
Focusing on the 100m reveals intriguing trends, including the ascendancy of Jamaica as a sprint powerhouse 
and the elevation of South African and East Asian sprinters to the global stage, a direct challenge to 
the racialist paradigm.  In line with an in-depth analysis of the influences on elite runners, we build a 
regression model to predict 100m performance based on environmental and psychological factors.

\noindent Next, we direct our attention to 1500m, where the last two British champions have been                              
part of a European resurgence that hasn't been seen in decades. Examining three                             
different time periods, we identify a thirty year national slowdown (1989-2018).  Adapting 
our model to this time period reveals striking evidence that \textit{racial perception}                          
has greater impact on performance than \textit{racial physiology}.                                                    

\noindent Synthesizing these findings, we introduce a psychocultural hypothesis, positing that interactions 
between racial perceptions and social dynamics shape the global distribution of running performance. 
We contrast this hypothesis with the racialist paradigm and propose extending it beyond sport where it 
offers insight across many domains.


\end{abstract}


\maketitle

\section*{Introduction}

\noindent For as long as humans have participated in global athletics, observers have essentialized the 
characteristics necessary to become champion \cite{history}.  That these ideas have largely been  
advanced by theorists lacking deep familiarity with athletics underscores a recurring theme: many 
of these proclamations, such as the impossibility of the four-minute mile \cite{limit0,limit1,limit2}, 
or that 100 meter times can be extrapolated to predict when women overtake men \cite{ladies}, have failed 
to withstand scrutiny.  In the first half of the twentieth century, when racialist explanations for 
athletic endowment deriving explicitly from white supremacy gained traction \cite{hitler1,hitler2}, Dr. Montague 
Cobb was able to draw upon his expertise as a preeminent anthropologist and experience as an accomplished 
athlete to refute these ideas \cite{cobb_bio, cobb}.  Since then, despite explicit racial determinism 
falling out of favor and modern genetics challenging traditional skin color described groups 
\cite{myth, miller}, many of these ideas remain popular - albeit updated to reflect modern scientific parlance. 
Most modern racialist theories focus on performance of specific "\emph{subraces}" (i.e. different regions 
in Africa) and obtain legitimacy through repeated assurances that vindication from improving genetic 
technology is \emph{just around the corner} \cite{gene1,gene2,peering,larsen,taboo,harp}. 

Writing for both the athletic and scientific community on the eve of the Barcelona Olympics (1992), 
Burfoot \cite{burfoot1,burfoot2} popularized the racialist lens by proposing that absent racial 
differentiation, we \emph{should} expect that Olympic medals be equally distributed across continents, 
and that absence of uniformity is evidence that West African's are uniquely endowed to sprint while  
East and South Africans have genetic advantages over longer distances.  Burfoot also proclaimed that 
over the next ten years scientists will: 

\begin{quoting} 
\emph{"decipher all 100,000 human genes, cure certain 
inherited diseases, and tell us more about ourselves than we are prepared to know, including, 
in all likelihood, why some people run faster than others}.
\end{quoting}

\noindent Twelve years later, writing for a general audience on the eve of the Athens Olympics (2004), 
Holden \cite{peering} supported Burfoot's hypothesis, by first reporting that altitude, diet, and effort had 
all been ruled out as explanations for Kenyan running superiority. Then, despite interviewing 
multiple scientists who reported that genetic evidence had \emph{not yet} been found to support the 
hypothesis, Holden concluded that not only are West Africans better suited for speed than whites but that: 

\begin{quoting} \emph{"the differences between East and West Africans are even more striking"}, that they 
\emph{"doubtlessly have a strong genetic component"}, and that,
\emph{"Next Month's Olympic games in Athen's should demonstrate yet again that 
    West African runners are built for speed and Kenyans built to endure"} \cite{peering}.  
\end{quoting}

\noindent A year later, writing for a scientific audience, Pitsiladis and Scott provided evidence in the 
form of a table that assigned "Black African Ancestry" to every 2004-era male record holder and subdivided them into a West African 
group that excels in sprints and an East or North African group that excels over long distances \cite{main}.  
Taking a more measured approach, Pitsiladis and Scott predicted that even with the most optimistic outlook that 

\begin{quoting} 
\emph{"genes considered important in extreme phenotypes -eg, sprint, power, strength, and endurance- and the way that they 
interact with other genes and with the environment to create an elite athlete will not be fully understood for many
years"}, \end{quoting} 

\noindent and expressed concern that gene doping might gain in popularity regardless of its effectiveness \cite{main}.

\FloatBarrier \begin{figure}[!h] \centering 
\includegraphics[scale=0.3]{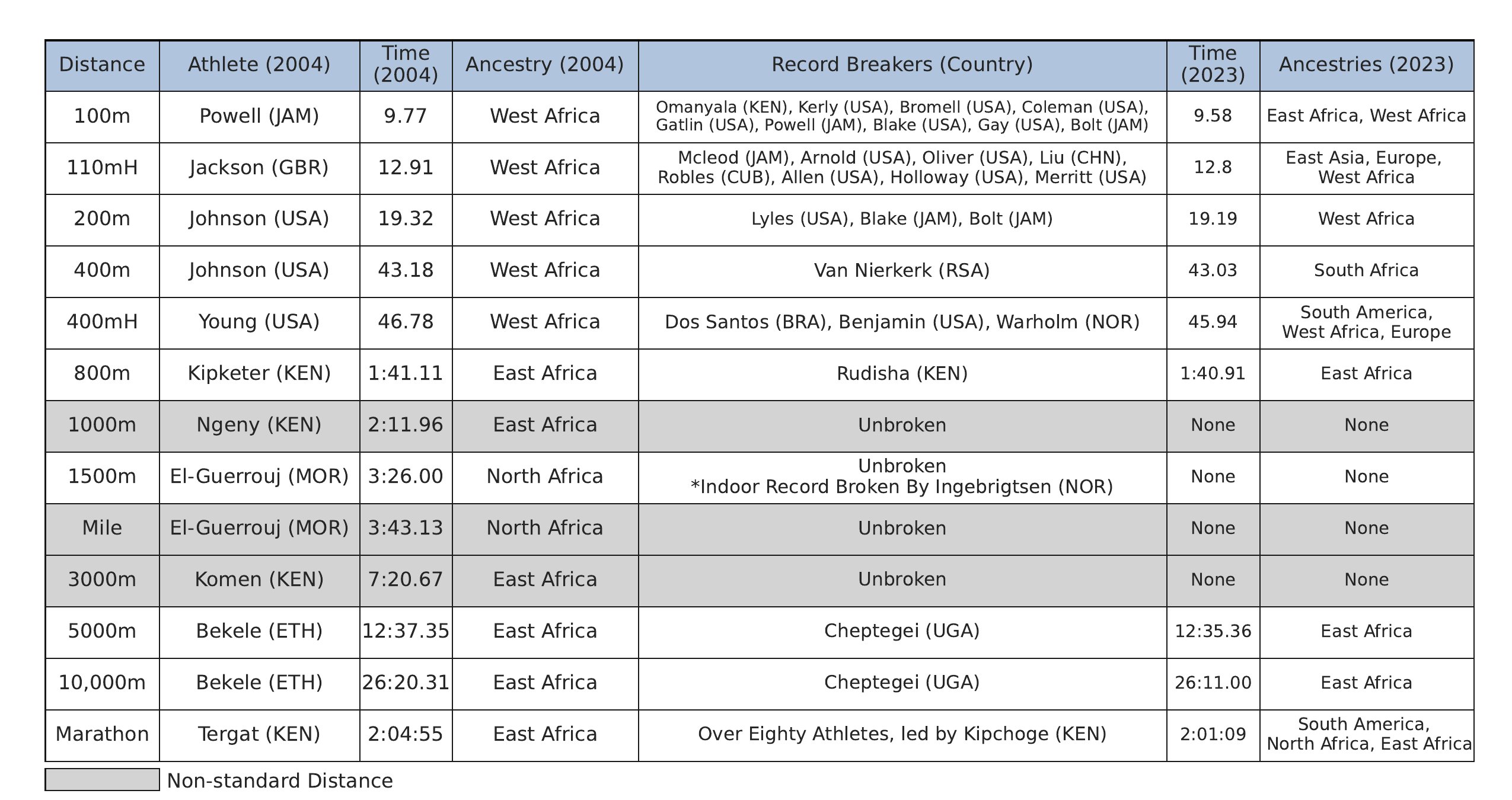} 
\caption{\footnotesize{\textbf{Men's World Running Records (Updated from 2004\cite{main})}}} 
\label{fig:intro} \end{figure} \FloatBarrier

\noindent In \fref{fig:intro}, we update the table supplied by Pitsiladis and Scott \cite{main} with 
every record broken and ratified as of July 2023. Several interesting patterns emerge. First, that while 
the assignment of every 2004-era record holder to a single genetic group - "Black African Ancestry" - was 
questionable (Ethiopians, Moroccans, and African-Americans are characterized by recent european admixture
\cite{emix, mmix, amix}), it is certainly no longer valid: Norwegian (400mH) and South African 
(400m, coloured ethnicity) athletes currently hold records and Chinese (110mH) and Brazilian (400mH) 
athletes have eclipsed the 2004-era records.  Focusing on specific events, we observe over 
eighty athletes besting the 2004-era marathon record, suggesting that technological 
advances \cite{shoes} have prevented performance saturation and made the event inappropriate for analysis.  
Excluding the marathon, the 100m mark has been surpassed by the most athletes (9) and every sprint event 
except the 200m has been surpassed by athletes without West African ancestry.  
In the middle distances, records from 1000-3000m still stand, and only one athlete has beaten the 800m, 5k, 
and 10k marks. However, recent shifts in the nationality of global championship winners who are quickly 
approaching these records require more detailed analysis.  In the following sections we provide in-depth 
analysis of the current state of affairs (2004-present) in the 100m and 1500m and observe that while 
genetic models are vastly overrated and racial naturalist models obsolete, there is compelling evidence 
that racial narratives (racism) impart psychocultural impact that is sufficient to measurably affect 
running performance. 

\section*{100m} 
\begin{figure}[!ht] \centering  
\includegraphics[trim=0cm 0cm 0cm 0cm, clip, width=\linewidth]{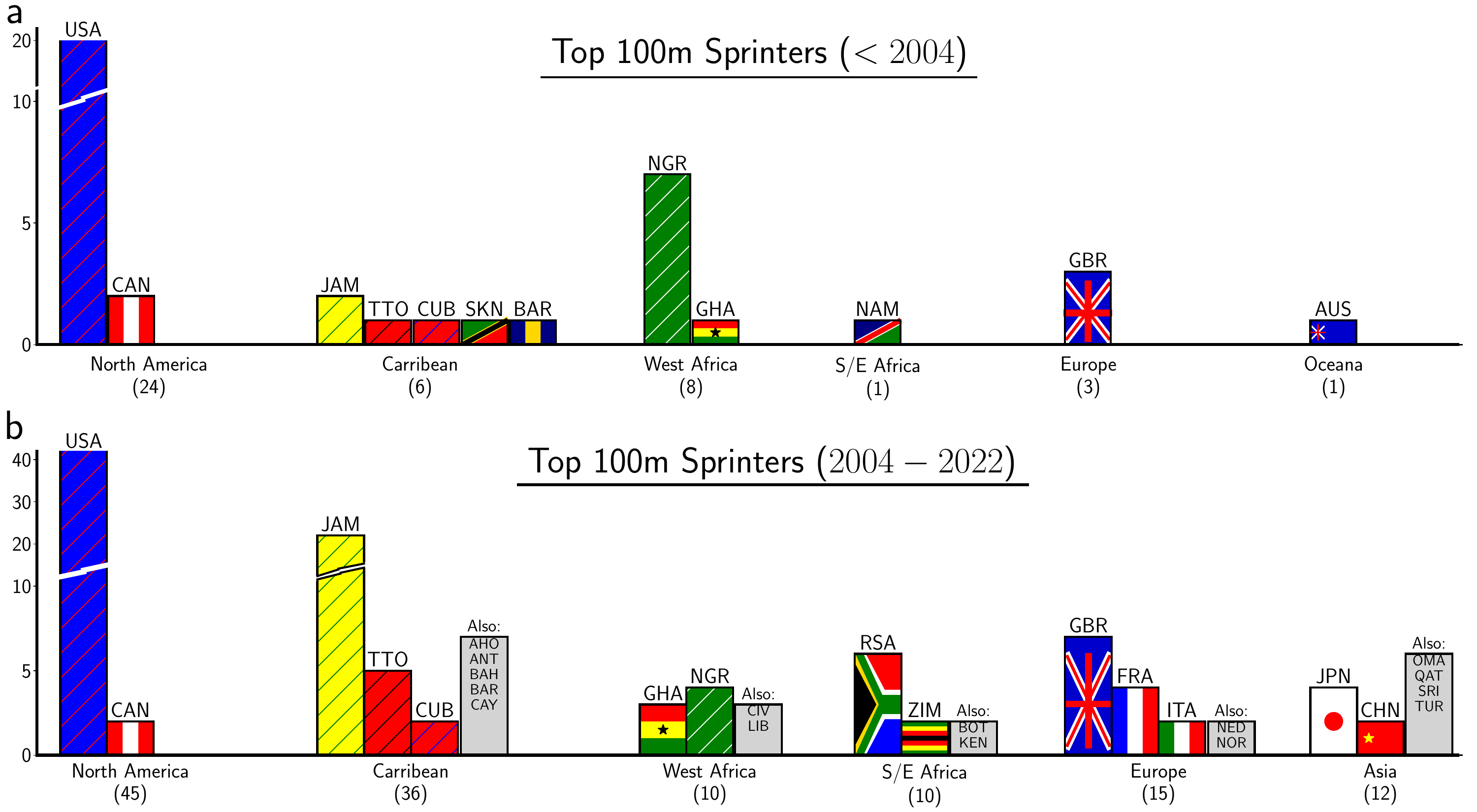} 
\caption{\footnotesize{
\textbf{a:} The distribution of elite (<10s) runners before 2004 supported the West African ancestry hypothesis.
\textbf{b:} The distribution of elite (<10s) runners since 2004 challenges the West African ancestry hypothesis. 
}}\label{fig:rel} \end{figure}

\vspace{2mm} 
\noindent Running 100m under ten seconds has long been considered a standard of excellence \cite{barrier}. 
Enumerating the nationality and geographic region of all athletes who have surpassed this 
mark provides a large sample that is robust to the small variations capable of determining world records \cite{windy}.  
In \fref{fig:rel}{-a} we observe the data ($<2004$) that motivated the West African hypothesis - 95\% 
athletes (excluding Patrick Johnson - an Australian of Aboriginal and Irish Ancestry, 
and Frankie Fredericks - Namibia) could plausibly be classified as West African.  Furthermore, in 2004, 
after the United States, the next most represented country was Nigeria which is the largest country in 
West Africa.  Taken together, these observations support the racialist hypothesis that genetic adaptation 
specific to West African ancestry provides significant advantages in sprinting.

\FloatBarrier  \FloatBarrier

\noindent In \fref{fig:rel}{-b}, we observe the distribution of nationalities for athletes who have eclipsed 
the 10-second standard after 2004. While the United States again leads the pack, the rise of Jamaica as a 
100m powerhouse is evident (a 10x increase in athletes), along with the emergence of athletes from East Asia 
and South Africa. One complication with this data is that, unlike in 2004, the North American and European 
lists now include multiple athletes of diverse ancestry, such as Gemili (North African Ancestry, 
representing Great Britain), and Boling (USA), Tortu (Italy), and Lemaitre (France), all who have no known 
recent African ancestry. Looking further at the modern 100m data we find four pieces of evidence that 
complicate or directly challenge the West African sprinting hypothesis:  

\subsubsection*{1. Caribbean Enrichment} ~\\ \vspace{-7mm} 

\noindent Since 2004, approximately 30\% of worldwide elite ($<10s$) 100m athletes have represented Caribbean 
countries and 20\% have been Jamaican nationals. Given that Jamaica accounts for less than 0.03\% (2.8M) 
of the global population, this represents a relative enrichment that is almost a hundred times greater 
than that for athletes classified by West African ancestry.  Already impressive, it's likely 
an \emph{underestimate} if one considers how many European and North-American athlete are directly tied 
to the Caribbean. For example, using birthplace rather than nationality, we find Jamaicans winning six of 
the last nine Olympic 100m finals. Extending this definition to include children born to parents of West 
Indian origin (e.g., Andre De Grasse, a six-time Olympic medalist from Canada has parents 
from Trinidad and the Bahamas) and the enrichment grows significantly.  Go even further 
and assign Jamaican identity using a single parent and the Nigerian 100m record (9.85s), 
which when set in 2006 made Olusoji Fasuba the seventh fastest man in the world, also belongs to Jamaica 
by way of his mother, herself a former sprinter and cousin to the famous Jamaican sprinter, Don Quarrie \cite{ola}. 

Should the West African explanation be discarded for a story about recent genetic adaptations that has 
resulted in Caribbean specific "speed genes"?  While certainly the more data driven hypothesis, it should be 
pointed out that modern genetics has allowed us to ask this very question and that the results have not 
been fruitful \cite{fail1, fail2}.  Moreover, looking further at Caribbean enrichment in athletics reveals
considerable island specific athletic success. For example, while Jamaica and Trinidad enjoy outsized success 
over 100m and 200m, Bahamas excels at sailing and 400m \cite{bahamas}, Grenada at the 400m and the 
javelin \cite{grenada}, and Cuba outperforms almost all nations in the technical jumps.  That different 
islands with different colonial histories, founding populations, and proportions of recent admixture 
\cite{wig}, might have each undergone event specific genetic selection is an exceedingly unlikely hypothesis. 
Additionally, a genetic argument from the gaps, overlooks multiple cultural and environmental factors, 
including but not limited to the substantial investment in Caribbean infrastructure \cite{dev}, coaching talent 
sufficient to draw in athletes from other countries \cite{ker}, and the massive social value placed upon 
Secondary School track and field competition in a region where cultivation of talent has historically been 
viewed as a pathway to a subsidized university education in the United States \cite{champs, champs2}. 
Without further evidence otherwise, we find no compelling reason to reject the parsimonious and well 
evidenced argument made by Jamaican author and retired sprinter Orville Taylor \cite{culture}, namely 
that \emph{it's culture, not genes} that describe these islands.

\subsubsection*{2. African Diversity} ~\\ \vspace{-7mm} 

\noindent Another pattern that has emerged over the previous two decades is the success of African sprinters 
from outside of West Africa, a phenomenon that is reflected in the following area and world records: 

\begin{itemize}
    \item The African 100m record is in East Africa (Omanyala - Kenya).
    \item The runner-up to the African 100m record is in South African (Simbine - RSA) 
    \item The 200m African (and World under 20) record over 100m is in South Africa (Tebogo - Botswana). 
    \item The 400m African (and World) record is in South Africa (Van Nierkerk - RSA, "coloured" or admixed ancestry). 
\end{itemize}

\noindent This pattern does not just reflect records, considering the top hundred 100m performances over the 
last five years by African athletes in the 100m, more have came from South Africa (34) and Kenya (20) 
than Nigeria (14), despite Nigeria having more than twice the combined population of South Africa and Kenya.  
These recent performances directly contradict one of Burfoot's central claims, that South
Africans are suited for long distance road-racing because 
\begin{quoting} 
\emph{"South African blacks are related to East African blacks through their common Bushmen ancestors.  
West African blacks, representing the Negroid race, stand apart"} \cite{burfoot2}.  
\end{quoting} 

\noindent Without passing judgment with respect to the anthropological validity of this claim, 
we conclude that it does not fit the data.

\subsubsection*{3. East Asian Success} ~\\ \vspace{-7mm}  

\noindent The third piece of evidence involves the emergence of Asian sprinters, 
twelve have ran under ten seconds since 2017.  This rise is bolstered by the following evidence: 
\begin{itemize}
    \item In the 4x100m relay, Japan and China together have earned medals in five of the last seven 
          global championships (2015-2023) after earning one in the previous 30 global championships (1950-2014).
    \item There have been more Chinese athletes (3) in the last ten global 100m finals than West African athletes (2). 
    \item The Chinese national record over 100m of 9.83 seconds is faster than any record in West Africa. 
          In setting this record, Su Bingtain recorded the fastest split over the first 60m of the 
          100m race in history.
    \item Two World Youth 100m records (<17 yrs and <18 yrs) are held by two different Asian athletes 
          (Kriyu - Japan  and Boonson - Thailand).
\end{itemize}

\noindent Again, we find that the data disagrees with the racialist hypothesis, namely the claim that, 
\begin{quoting} 
\emph{"when pure explosive power - this is, sprinting and jumping - is required for excellence in a sport, 
    blacks of West African heritage excel"} 
\cite{burfoot2}, 
\end{quoting} 

\noindent and reflect on the fact that last two World Championships in the long jump (an event that directly combines 
sprinting and jumping) were from China and Greece.  Not only does the emergence of 
Asian sprinters undermine stereotypes about race and masculinity \cite{cmen}, it also complicates genetic 
explanations because the average genetic distance between East Asian and West Africans populations 
is as far as any continental comparison \cite{adist}.

\subsubsection*{4. Morphological Diversity} ~\\ \vspace{-7mm} 

\noindent While the global distribution of elite 100m runners post-2004 clearly doesn't line up with the 
West African hypothesis, an equally compelling reason to discard the racialist argument is the morphological 
and biomechanical diversity observed across elite 100m athletes. Biomechanical analysis demonstrates multiple 
pathways to elite speed - some athletes execute long powerful ground contacts and exhibit strength levels 
comparable to weightlifters \cite{spring} while others rely primarily on elasticity and abstain from 
weightlifting completely \cite{collins}.  These differences are increasingly being recognized by elite 
speed coaches, whose recognition of these differences has led them to classify sprinters into models or types 
whose qualities are so divergent that optimal training for one athlete can be counterproductive for another.

After the 6'5" Usain Bolt became internationally recognized for setting records in 2008, 
scientists quickly explained why height is necessary for speed \cite{size} and showed why long ground 
contacts and change in center of gravity allows athletes to produce long strides necessary for top speed.  
However, these explanations hardly applied to the following year's NCAA champion, 5'5" Trindon Holliday 
\cite{trindon}, whose running model was based on rapid acceleration via explosive muscular strength and 
linearly diminishing ground contact times.  A more recent example of this race model is Su Bingtian of China, 
who at 5'7", has run the first 60\% of the race faster than anyone else in the world.  Importantly, this 
is not a Chinese model, the second fastest Chinese sprinter (Xie Zhenye, 6'2") uses the same high frequency 
stride and maintenance of center of mass that allows sprinters like Tyson Gay and Wade Van Neirkerk to excel 
at sprints from 100m to 400m.  While Bolt's biomechanics are unique - the athlete that experts have pegged 
as his greatest imitator is 6'4" Christophe Lemaitre \cite{lemmy} of France, who became the first European 
without recent African ancestry to break ten seconds just four years after he was discovered via a sports 
campaign to measure fifteen year olds over 50m \cite{lem2}.  While it may appear to outside 
observers that sprinting is a universal activity that provides a 
\emph{"perfect scientific laboratory for the exploration of physical and performance differences 
between racial groups"} \cite{burfoot2}, 
this couldn't be further from truth; at the elite level sprinting is a complex activity where different 
strategies and qualities trade-off \cite{comps} to determine the end result.  Just like horses, hares, 
greyhounds and ostriches achieve similar top speeds (70 kmh) through very different mechanisms, so to do 
humans at the elite levels.  Thus, the idea that a certain ancestry group might have a monopoly on the 
multitude of different qualities that can make a world-class sprinter is as akin to the idea that a 
single group of soccer playing people have been genetically selected to excel at midfielder and goalkeeper.

\section*{1500m}  
\vspace{2mm} 

\begin{figure}[!h] \centering 
\includegraphics[trim=0cm 0cm 0cm 0cm, clip, width=\linewidth]{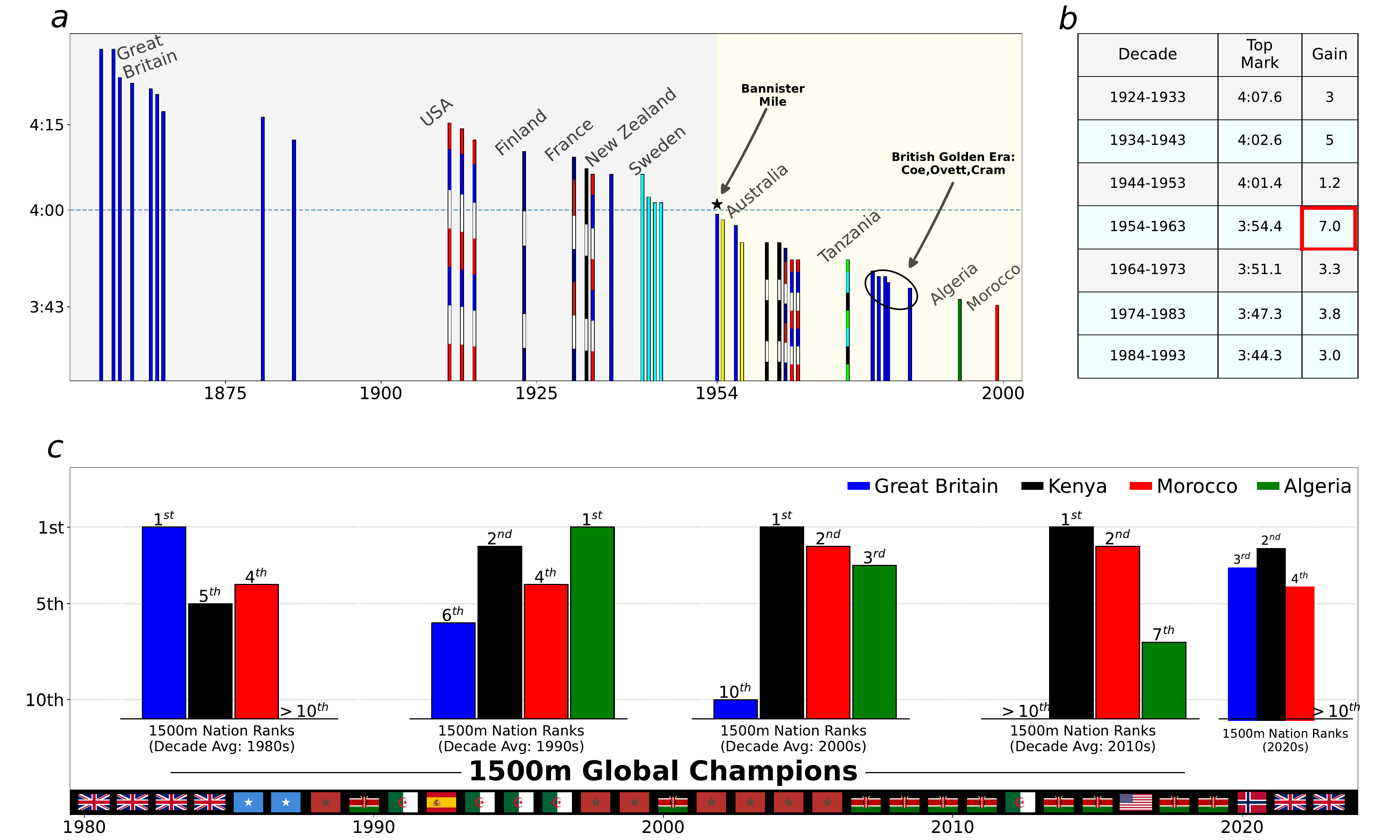}
\caption{\footnotesize{
\textbf{a:} Mile world record progression by country. 
\textbf{b:} The decade after Roger Bannister's 4-minute mile saw the largest gain (7 seconds). 
\textbf{c:} Average national rankings by country, by decade. Great Britain went from 1st to out of the top ten.  
}}\label{fig:mile} \end{figure}

\noindent The 1500m race is the only major championship event whose record was set in 
the previous century (El Gerrouj, 3:26:00, 1998). Often referred to as the \textit{metric mile} (a mile 
is just 7\% further) it holds particular significance in the Anglosphere where the 
longer distance has been contested since the middle of the nineteenth century (\fref{fig:mile}{-a}). 
When progress appeared to plateau after the second world war, scientists declared it may be 
physiologically impossible to break that "four minute barrier" \cite{ban2}, perhaps further 
delaying the event - which finally occurred in 1954 when Roger Bannister of England ran 3:59.4 \cite{bannister}. 
The "barrier" now broken, another sub four clocking occurred again only 46 days later, 
eight more times over the next three years, and the following decade saw the largest improvement since 
records were first recorded (see \fref{fig:mile}{-b}).  
Catalyzing an era of Anglosphere excellence, the next thirty years saw over half of 1500m global 
championships go to men from Commonwealth nations and a group of British runners dominate the middle distance 
events by the early 1980s. This excellence peaked in the 1984 Olympic 1500m final, when the British 
men who qualified for the final represented the defending Olympic champion (Seb Coe), 
the defending world champion (Steve Cram), and the current world record holder (Ovett 3:30.77).  

\FloatBarrier  \FloatBarrier 

\noindent Despite British athletes winning two medals (as they had in the 1980 Olympics) and 
picking up one more four years later, tides were changing.  Over the next 30 years 95\% (21 of 22) 
of global 1500m championships went to men born in in Africa - after the 1988 silver medal, 
British men didn't see the podium again for over thirty years.  The average annual rank of the leading 
British 1500m time (compared to other countries) went from first (1980s) to sixth (1990s) and then 
to outside of the top ten in the 2000s (see \fref{fig:mile}{-c}).

This rapid changing of the guard led many to surmise that genetic advantages specific to African 
ancestry \emph{must} exist and spurred research to explain how they manifest \cite{north,marathon,lance}.  
Closely examining this hypothesis reveals some complications.  Firstly, the idea of a genetic group 
includes Algerians, Moroccans, and Kenyans but excludes West Africans and Europeans appears motivated 
almost exclusively by shared athletic performance rather than ancestry or geography.  In this case, 
North Africans are treated as "Schrodinger's Africans": racialized as \emph{Black Africans} when 
running \cite{main,stt,peering,lance} but as \textit{White Caucasians} when the discussion shifts to the 
number of African chess grandmasters and the "Sub-Saharan distinction" becomes critical \cite{chess1,chess2}.  

Perhaps because of this complication, the majority of candidate gene studies have focused on Kenyan and 
Ethiopian runners, and as was the case with Jamaican sprinting genes, efforts to locate the genetic link 
have been underwhelming \cite{null,k_null1,k_null2,k_null3}.  Interestingly, what has been discovered in 
studies of mtDNA and Y chromosome haplotypes was that neither Kenyan nor Ethiopian elite athletes are 
genetically distinct populations and that a significant fraction of both elite athletes and controls share 
recent maternal and paternal ancestors with Europeans, suggesting that "race" and "running" are unlinked 
in these populations \cite{k_funny1,k_funny2,k_funny3}.  With these studies in mind we query the last 
forty years of 1500m performance data and find two crucial pieces of evidence that directly contradict 
the theory of racial genetic specialization.

\subsubsection*{1. The British Have Slowed Down} ~\\ \vspace{-7mm}

\begin{figure}[!h] \centering 
\includegraphics[trim=0cm 0cm 0cm 0cm, clip, width=\linewidth]{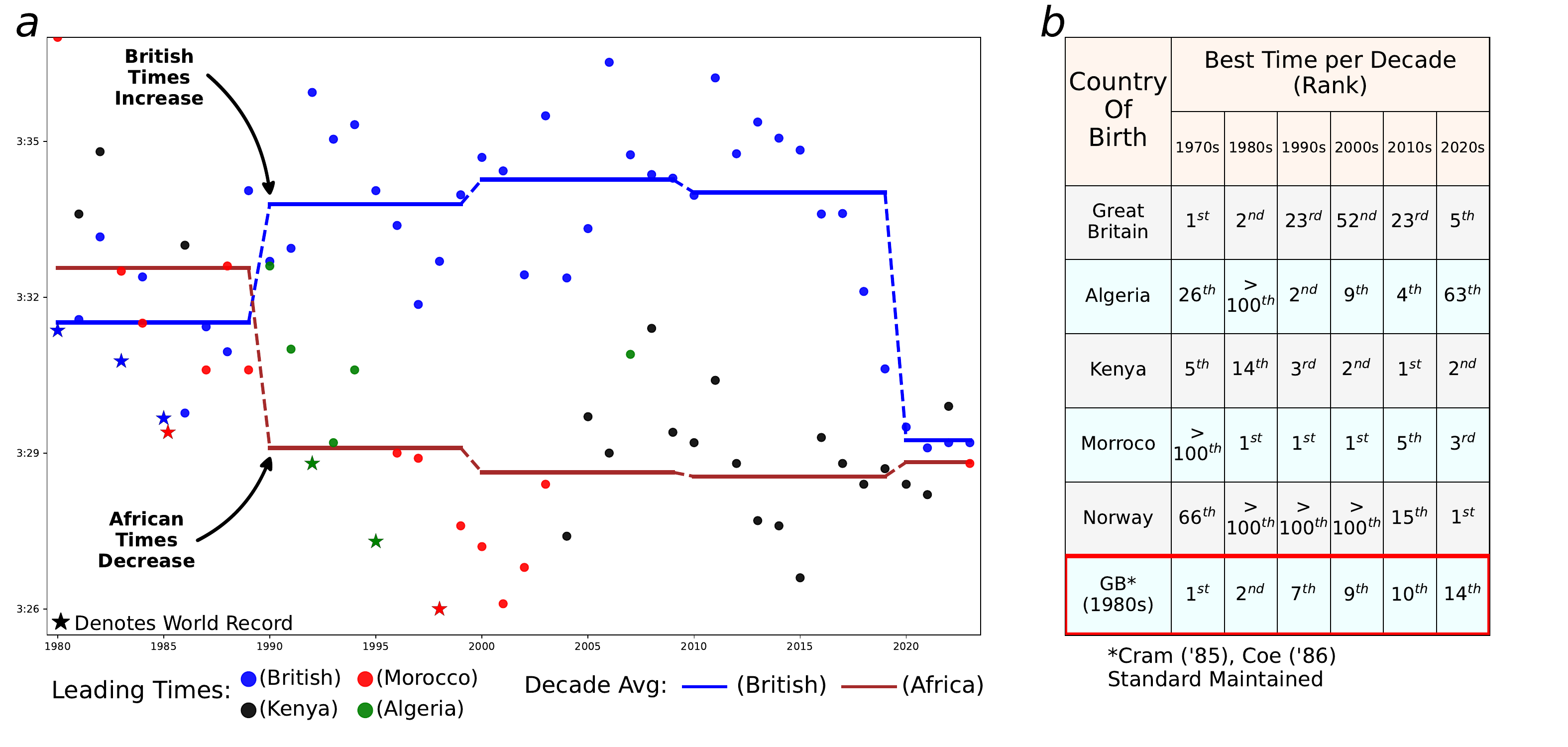} 
\caption{\footnotesize{
\textbf{a:} Starting in the 1990s British times have gotten slower while African times have gotten faster.  
\textbf{b:} Average individual rankings by country, by decade. Top British athletes average rank was 
went from 3rd, 38th, 70th, 60th, 5th from the 1980s to 2020s.  In the bottom row hypothetical rankings 
are shown that assumes Britain maintained one of their top two times from the 1980s. 
}}\label{fig:slowing} \end{figure}

\noindent It is well known that a performance gap opened up between European and African 
born middle-distance athletes coinciding with the end of British dominance over the distances 
\cite{not_slower}. What is less well known is that a large proportion of this gap was caused by British 
athletes slowing down.  In \fref{fig:slowing}{-a} we show that despite continued improvements to track 
surfaces and footwear \cite{tech1,tech2}, annual leading British 1500m times slowed significantly from the 
1980s to the 1990s (3:31.5 to 3:33.8, p-value $<$ 0.0015) and stayed this way for approximately thirty years.  
This drop-off is not confined to the 1500m.  To this day every other British-born middle distance 
record (800m, 1000m, the mile, 3000m, and 5000m) was set between 1981 and 1986.  In \fref{fig:slowing}{-b} 
athletes best times per decade are ranked by nationality and the British move from 1st and 2nd to 
23rd and 52nd from the 1970s and 1980s to the 1990s and the 2000s (top row) .  
In the bottom row we see that if the British had maintained the standard set by Cram ('85) or Coe ('86), 
that they never would have fallen out of the top ten, despite being decades behind with respect to footwear 
and training technology - something hard to square with the claim that genetic differences are behind 
the disappearance of the British miler \cite{rulez}. 

\FloatBarrier  \FloatBarrier

\subsubsection*{2. The British Have Sped Up} ~\\ \vspace{-7mm} 

\noindent If genetics cannot explain the thirty year British slowdown (1989-2018), then genetics certainly 
cannot explain the recent reemergence of British athletes 1500m athletes in the last five years. 
Comparing these eras we find British athletes running the 1500m under 3:32 \textbf{(1)} twelve times in 
the eight years from (1980-1988), \textbf{(2)} once in next thirty years (1989-2018), and \textbf{(3)} 
fifteen times in last five years (2019-2023).  After never coming within two seconds of Cram's 1985 mark 
(1989-2018), they have surpassed it five times in the last five years. 
Perhaps more importantly, 
they have excelled in crucial moments, earning their first Olympic 1500m medal in over thirty years in 2021 
and winning last two world championships (2022 - Wightman and 2023 - Kerr), part of a return to glory 
that should make even the most committed hereditarians question racialist explanations for 
middle distance excellence.  

\vspace{2mm} 
\section*{The Psychocultural Hypothesis} 
\vspace{2mm} 

\noindent Our analysis has so far focused on whether racialist explanations of athletic success stand up 
to scrutiny.  Analyzing the success of sprinters outside of West Africa, the morphological and biomechanical 
diversity associated with 100m success, and the return of the British middle distance runner, a compelling 
case can be made against racial determinism.   What we have not done is attempt to explain \emph{why} the 
disparities that are often taken as a priori evidence of racial specialization \cite{burfoot2} exist in 
the first place.  Here we develop a framework that better explains why non-uniformity is the norm with 
respect to the global distribution of runners and provides answers to the question: \emph{if not race, then what?}  

Previously, in our analysis of Caribbean sprinters, we posited that enduring cultural and national 
interests can go a long way toward explaining the islands affinity for sprinting.  However, under any 
definition, "national culture" is insufficient to explain the rise, fall, and rise of British milers. 
Dynamic changes in performance at the national level cannot be explained by genetics or cultural values - 
a more immediate impetus acting on the level of athletes and coaches is required.  This perceptual force, 
interacting within a local cultural milieu is the \emph{psycho} in "psychoculturalism".  To develop this 
idea we first turn to individual event data - keeping our focus on the 100m and 1500m - to identify and 
quantify sources of psychological impact on the competitors.

\FloatBarrier \begin{figure}[!h] \centering 
\includegraphics[trim=0cm 0cm 0cm 0cm, clip, width=\linewidth]{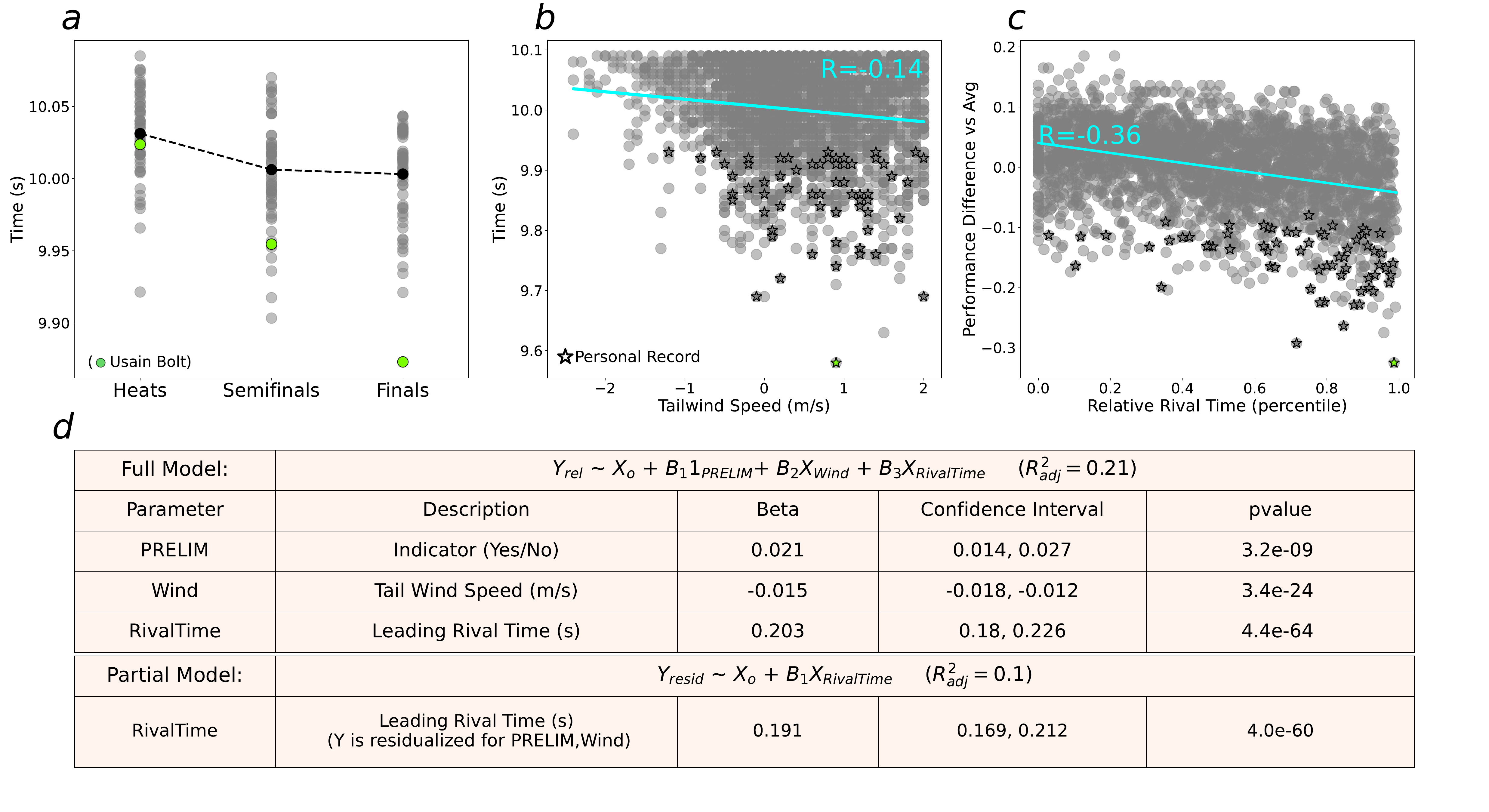}
\caption{\footnotesize{
\textbf{a:} Athletes run the heats in championship meets slower to preserve energy.  With the exception of Usain Bolt, 
athletes do not run semi-final races measurably slower than the finals. 
\textbf{b:} Greater tailwind speed is associated with faster times. 
\textbf{c:} The time of the fastest competitor is associated with faster times. 
\textbf{d:} In the full model: PRELIM, Wind, and RivalTime are all significant variables with respect to prediction
    of $Y_{rel}$, the difference an athletes time is from their average.  In the partial model, RivalTime is still
    significant after on after $Y_{rel}$ is residualized for PRELIM and Wind. 
}}\label{fig:comp} \end{figure} \FloatBarrier

\subsubsection*{1. Foes as Friends over 100m} ~\\ \vspace{-7mm} 

\noindent As shown in \fref{fig:comp}, the 100m performance of elite athletes 
(PR < 9.95 seconds, 10 races under 10.1 seconds) can be modeled using three variables, 
\textbf{(a)} whether the race occurs in a heat, a semi-final, or a final (or invitational),
\textbf{(b)} windspeed, previously reported to be associated with faster performances \cite{windy,windy2}, 
and \textbf{(c)}, the time of the fastest rival in the race.  In testing this model we find that: 
\begin{itemize}
    \item Times are significantly slower in the heats but not significantly faster in finals vs semifinals. 
    \item Greater tailwinds are associated with faster time. 
    \item Rival time is associated with faster times in the full model and on the residuals from tailwind and PRELIM. 
\end{itemize} 

\noindent Scaling effect sizes, the average benefit to an athlete from a rival running an additional 0.1 
seconds faster is greater than the boost from an additional 1 m/s of tailwind.  Hundred meter athletes 
run in separate lanes and do not significantly influence drag meaning that this benefit  is more likely to 
be psychological than biophysical.  Noting that differences between elite athletes are quite small and that 
there are undoubtedly diminishing returns to this effect suggests that in the range where it is experienced 
- rivals in very close contact - it may be of even larger benefit than predicted by the regression model, 
corroborating previous research suggesting that competitors may unintentionally synchronize their 
strides to match their opponents \cite{sync_yes,sync_no}.

\subsubsection*{2. Middle Distance Inspiration} ~\\ \vspace{-7mm}

\begin{figure}[!h] \centering 
\includegraphics[trim=0cm 0cm 0cm 0cm, clip, width=\linewidth]{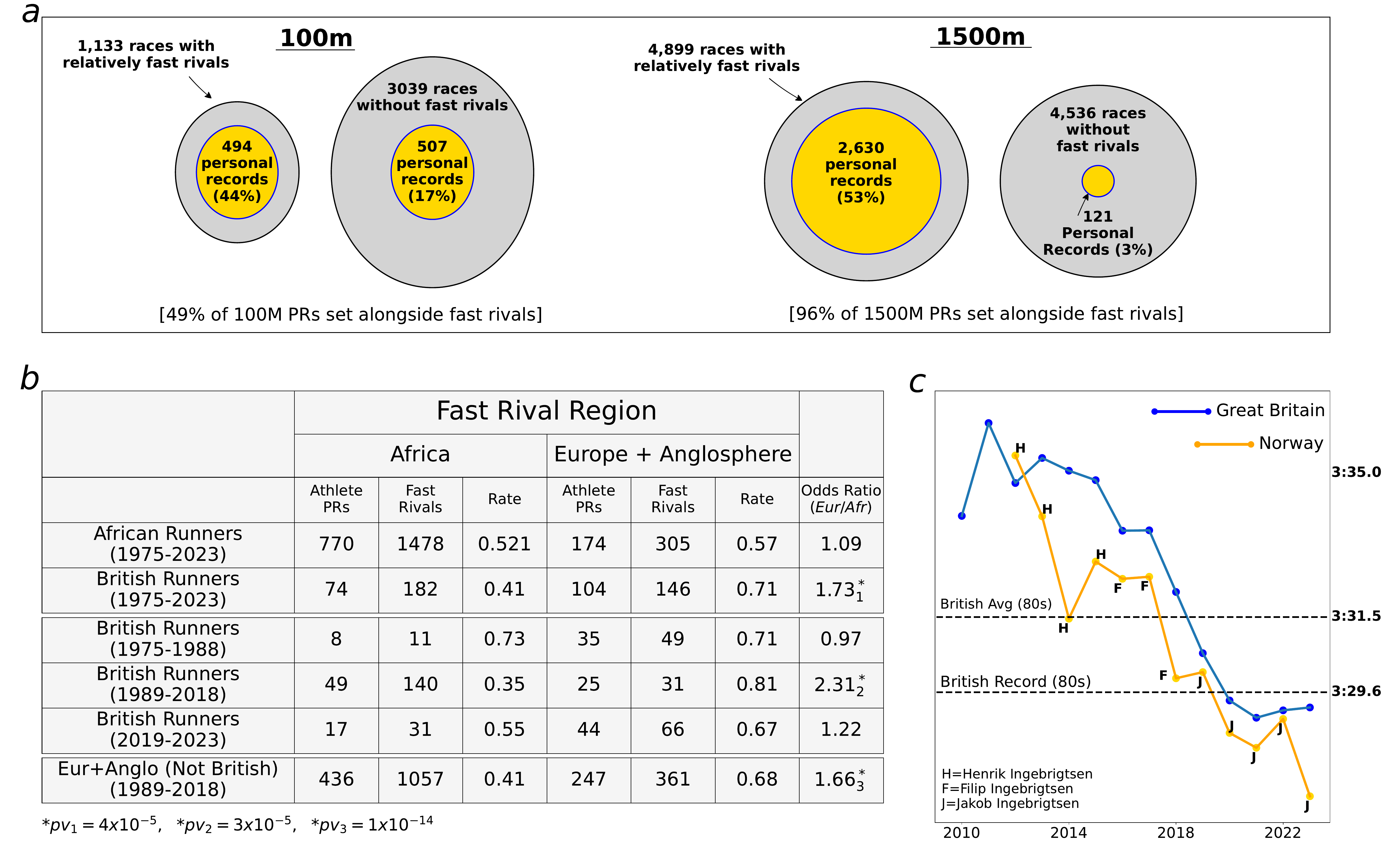}
\caption{\footnotesize{
From the standpoint of runner $X$ whose personal record before the race is $t$, a fast rival is present 
in the race if someone other than $X$ runs $<t$.  If $X$ wins the race then the fast rival is the $2^{nd}$ 
place finisher, otherwise the fast rival is the winner.  
\textbf{a:} Records are more likely to be set alongside fast rivals, especially for the 1500m 
\textbf{b:} African and British runners respond differently depending on the nationality of rivals. 
\textbf{c:} British improvement tracks with Norwegian performance. 
}}\label{fig:brit} \end{figure}

\noindent In middle distance events the benefits of a speedy lead runner are so well established that athletes 
often employ "rabbits" (pacesetters) to lead a large portion of the race before dropping out.  Because distance 
events take place in single lane there is some speculation that rabbits provide a biophysical 
benefit (reducing air resistance), however most analyses have estimated this effect to be almost 
negligible, especially in the 1500m \cite{drafting}.   In major championship races, where official 
pacesetting is prohibited, athletes frequently coerce competitors into unwittingly playing the role of 
the rabbit \cite{1500m_tactics1,1500m_tactics2}. The end result of this in-race psychological warfare is 
that runners are more than fifteen times more likely to break a personal record in races where they are 
not the only athlete to go under their previous best (see \fref{fig:brit}{-a}). 

\FloatBarrier  \FloatBarrier 

\noindent Having established that competitors provides real non-physical benefits, we explore whether it is 
only related to psychological fatigue (i.e. not concerning oneself with the pace preserves energy) 
or whether rivals provide a measurable inspirational component (i.e. seeing is believing, even in real time).
We surmise that if the latter plays a significant role that it may manifest similar to the way that 
higher-profile events are thought to inspire non-participating athletes, and recall Roger Bannister's 
four minute mile and his prophetic post-race quote: 

\begin{quoting} 
\emph{"although physiology may indicate respiratory 
and circulatory limits to muscular effort, psychological and other factors beyond the ken of 
physiology set the razor’s edge of defeat or victory and determine how close an athlete 
approaches the absolute limits of performance”} \cite{ban3}. 
\end{quoting}

\noindent Assuming the quote had outsized impact on athletes, coaches, and the broader 
public in the Anglosphere \cite{banny}, we ask: \emph{If we assign some credit to Bannister 
mile for inspiring the generation of British runners that dominated the middle distances in 1975-1985, 
might we also identify an event or series of events that sufficiently discouraged the following generation?} 
Beginning with the summer of 1985, we recall \emph{how} the British hold on middle distances was 
relinquished by considering the following timeline:

\begin{enumerate}
    \item[\textbf{July 16:}] Steve Cram sets the 1500m record in classic fashion in Nice, France, building a large lead and barely surviving 
                                             a furious finish from Said Aouita, an emerging Moroccan talent who also eclipses the previous record. 
    \item[\textbf{July 22:}] A few days later in Oslo Aouita takes the 5000m record from the great British runner, David Moorcroft. 
    \item[\textbf{Aug 21:}] A month later in Zurich, Aouita suffers a hamstring injury and barely misses Cram's mile record. 
    \item[\textbf{Aug 23:}] Not one to be denied for long, Aiouta travels to Berlin two days later and takes down the 1500m record. 
\end{enumerate}

\noindent Hailed as a "special talent", analysts marveled at Aouita's ability to combine "natural speed" 
with endurance and credited him with inspiring the next generation of North African runners including 
the next two athletes to break the record, 
Morceli of Algeria \cite{idol} and El Geurrouj of Morroco \cite{idol2}.  Noting that this role as 
a singular inspiration might be exaggerated for geographic convenience - Morceli patterned his running 
after the Czech runner Emil Zapotek \cite{n_book,n_book2} and El Geurrouj after the Finnish athlete
Paavo Nurmi \cite{g_book}, we still consider the corollary; if Aouita can be credited with ushering 
in change and inspiring an entire continent, might he also have played a role in demoralizing European 
athletes?   And if not Aouita himself, might the racialist narrative that 
"\emph{White Men Can't Run}" \cite{stt}  that grew in popularity and perceived scientific 
legitimacy \cite{stt2,s_racism} have had an impact on running or coaching psychology? 

There is evidence to support this claim (see \fref{fig:brit}{-b}).  While all 1500m athletes benefit from 
competition and are more likely to set a personal bests alongside fast rivals, the benefits are not uniformly 
distributed.  British born athletes are more likely to benefit from a non-African born rival (1975-2023, 
odds-ratio 1.73, pvalue < 0.0005) a difference that is driven by primarily by the thirty year slowdown 
from 1989-2018, a time period where British-born athletes were more than twice as likely to break a personal 
record alongside a non-African rival (odds ratio 2.31, pvalue < 0.00005).  Interestingly, we see that this 
unfortunate disparity (British athletes had 4.5 times as many opportunities alongside African runners from 
1989-2018, 140 versus 31), has all but vanished during the British return to glory (2019-2023, odds ratio 1.22, pvalue > 0.5).

If we attribute Bannister's mile as the catalyzing force behind "Track Brittanica" (1975-1988), and explain 
the slowdown associated with the "Post-Coe Blues" (1989-2018) as a product of the negative racial perception 
that manifested as a popular response to African runners, then to what can we attribute the 
"Return to Glory" (2019-2023)?  One potential clue is illustrated in \fref{fig:brit}{-c}; since 2013 British 
annual leading times have tracked down almost perfectly in response to the annual leading times from Norway, 
all of which belong to three brothers, Henrik, Filip, and Jakob Ingebrigsten.  That three siblings have each 
taken a turn as their nations fastest runner could be interpreted as evidence of genetic evidence.  However, 
given the failure of studies to identify highly penetrant variants \cite{fail3} and the fact that 
heritable polygenic traits have half their variance within families \cite{barton2002,barton2005,t_sibs}, 
this familial accomplishment is more likely to signal shared environment even if an underlying genetic basis exists. 

Jakob, the youngest of the Ingebrigtsen brothers, has recently set world records in the 2000m, the 2-mile run, 
and the indoor 1500m.  He is known not only for his superb talent but for his brash personality and the 
viking heritage that he shares with countryman and current 400mH world record holder, Warholm 
(see \fref{fig:intro}).  Explaining his refusal to be intimidated: 

\begin{quoting} \emph{"There’s always been a thing in running where 
nobody can beat the Africans, but they have done what white people are afraid to do.
They start at an early age and commit to running from when they are five or six, 
because they don’t have any choice"} \cite{jg2} \end{quoting}

\noindent Jakob has ran in seven races with fast rivals from Africa and lowered his time in four of them.  
While it is unlikely he will get this chance again (the only runners ahead of him are retired), it is 
undeniable that he has played a direct role in the British return to glory (2019-2023).  The winner of the 
first two races where British born athletes first eclipsed Steve Cram's 1985 1500m mark, Ingebrigtsen 
has ran in a total of 27 professional races with British athletes.  In those races, British athletes have ran 
under 3:30 five times and under 3:31 nine times.   Without Ingebrigtsen, over thousands of races, 
British athletes have ran under 3:30 twice and under 3:31 10 times.  It isn't just fast times, 
Ingebrigtsen has also featured in each of the last three major championships.  He won the 2021 
Olympic 1500m final where Josh Kerr became the first British medalist in over 30 years.  
He was also runner up in the 2022 and 2023 World Championship final - in both cases bring passed by a British runner 
in the final part of race - events that have led to a very contentious rivalry \cite{kerr1,kerr2} 
that is pushing all involved to new heights.

\section*{Summary} \label{sec:discussion}

\noindent In this manuscript we have analyzed the last twenty years of race data for the 100m and 1500m and found 
it in conflict with the dominant racialist models that suggest regional African ancestry has a role in 
determining running ability. With respect to the 100m, the emergence of South African and East Asian sprinters 
challenge the argument that West African ancestry rather than culture is what underlies Caribbean success in sprinting.  
Cataloguing the morphological diversity and biomechanical variation across elite 100m sprinters, 
also calls into question whether it's even possible that a single ancestry group could have been 
selected for elite sprinting ability in the first place - a reason not to be surprised that attempts to locate "speed genes" have largely failed.
While arguing that genetic ancestry has been overstated as an influence on performance, we show that the 
impact that competitors have on each other is largely underappreciated and build a regression that shows that 
above and beyond race magnitude and environmental conditions, iron sharpens iron and does so during the race. 

We propose a psycho-cultural framework that considers multiple sources of psychological impact that occur 
within and between races. This framework allows us to quantify and validate what many athletes have 
frequently expressed, that belief is paramount.  Among many athletes who 
have been constrained by such beliefs, is the African 100m record holder Ferdidand Omanyala, who has has 
stated that:

\begin{quoting}"\emph{most of the people don't believe Kenyans can sprint but we are changing that 
slowly...It's just in the mind, change the mind and everything changes"} \cite{kenny}.\end{quoting}

\noindent Su Bingtian, the Asian 100m record holder, has expressed similar ideas, admitting that he previously 
didn't believe it was possible for him to run as fast as he has \cite{benny}, and explaining that: 

\begin{quoting} \emph{"Ten years ago, if you were going to claim that Asians can run under 10 seconds, many experts would immediately object to how impossible, 
drawing from whatever anatomical findings or racial theories out there.  Now if you can claim that Asians can one day break Bolt's record, 
indeed anybody, no matter whom, should accept any limit as given"} \cite{benny2}. \end{quoting}

\noindent Moving to the 1500m we applied our psycho-cultural lens to describe three periods in British middle 
distance running that we labeled "Track Brittanica" (1975-1988), the "Post-Coe Blues" (1989-2018), and 
"Return to Glory" (2019-2023).  We have shown that \emph{racial genetics} cannot possibly explain 
performance changes occurring over fewer than fifty years but \emph{racial perception} absolutely can.  
Building upon the 100m competition model we show systemic British underperformance among African runners 
co-occurring with the slowdown that is frequently mis-attributed to racial genetics.  Next we show that "Return to Glory" tracks with the emergence of a 
Jakob Ingebrigtsen, a Norwegian athlete who has explained his explicit rejection of racial narratives:

\begin{quoting} 
\emph{"I do not believe there is anything genetic that makes the Africans better. People use it as an excuse. 
It is why I work so hard and give up so much. I want to be the best in the world, not Europe, and so I train and
    live like this"} \cite{jg1}.
\end{quoting} 

\noindent Note that this result contradicts a common assumption - that the negative impacts of stereotypes \cite{st} 
and the need for role models that \emph{"look like you"} are concepts that apply only to those who are 
racialized as "non-white".  In this case the opposite is true, African runners who frequently emulate 
role models from other continents and are equally likely to be motivated running alongside anyone else 
have their accomplishments diminished while European and Anglosphere athletes find themselves 
unable to be motivated when not competing with co-ethnics (see \fref{fig:brit}{-b}).  This scenario 
illustrates why even so called "positive racial stereotypes" are so corrosive - diminishing the accomplishments 
of those they apply to and providing excuses for others does nobody any favors.  Thus, appealing to 
"racism of the gaps" to make genetic conjectures is self-serving - while these proclamations are all but 
guaranteed not to stand up to the test of time - repeatedly making them also extends their lifespan. 

Finally, while the authors of this paper are especially interested in Track and Field, the psychocultural 
framework that we have developed is by no means exclusive to sports.  The influences on performance that we 
have identified in here likely manifest across a multitude of human endeavors.  As an example, we consider 
that commonality between running and video-game playing: that both are attempted by many and pursued 
professionally by very few, and observe the recent developments in Tetris a classic video game known for 
being simple, yet cognitively challenging \cite{tetris_cog}.  Despite being out since 1989, and maintaining 
a large enough cult following to support yearly championships since 2010, it was finally completed for 
the first time at the beginning of 2024 \cite{tetris}.  And then twice more in the next 
month \cite{tetris2} - suggesting that even without the strictures of scientific racism, limits are 
subconsciously set far below the pinnacle of human achievement across many different domains.  For this 
reason we believe that there is so much progress to be made by not only rejecting the limits imposed 
by racism but by becoming universally motivated and allowing ourselves to be inspired by every other human, 
with no regard to whether they "look like us" or even whether they already exist or we have to become them.

\appendix
\section*{Appendix}

\printcredits

\bibliographystyle{model1-num-names} 

\bibliography{refs}


\end{document}